\pdfoutput=1
\documentclass{appolb}
\usepackage{epsfig}
\usepackage{float}
\usepackage{amsmath}
\def\Journal#1#2#3#4{{#1} {\bf #2}, #3 (#4)}

\def\NPB{{\em Nucl. Phys.} B}

\def\PRL{\em Phys. Rev. Lett.}
\def\PRD{{\em Phys. Rev.} D}

\def\MET{$\not$E$_t$~}

\begin{document}
\title{Tevatron collider program - physics, results, future? %
\thanks{Presented at Cracow Epiphany Conference on The First Year of the LHC, January 10-12, 2011, Krakow, Poland}%
}
\author{Krzysztof Sliwa
\address{Tufts University}
}
\maketitle
\begin{abstract}
An overview of more than 25 years of the Tevatron Collider program at Fermi National Accelerator Laboratory in Batavia, near Chicago, Illinois, USA, is presented. The physics goals of the program itself, the Tevatron accelerator design characteristics and some of its achievements are described. A selected set of the past and ongoing physics analyses and measurements performed by CDF and D0 collaborations are summarized. Also, in view of the modified  plans and schedule of the Large Hadron Collider (LHC) at CERN,  the future of the Tevatron program is discussed.
\end{abstract}
  
\section*{Standard Model (and beyond)}
The current (since $\sim$1975) understanding of elementary particles and their strong and electro-weak interactions in given by the Standard Model~\cite{SM}, a gauge theory based on the following symmetries:
\begin{eqnarray*}
SU(3)_c \times SU(2)_I \times U(1)_Y
\end{eqnarray*}
The $SU(3)$ is an unbroken symmetry of quantum-chromodynamics (QCD), a quantum theory of strong interactions whose carriers (gluons) are massless and couple to color (strong force charge). The $SU(2) \times U(1)$ is a gauge symmetry of the electroweak interactions. It is spontaneously broken by the Brout-Englert-Higgs mechanism, which gives masses to the electroweak bosons ($W^+,W^-, Z^o$ and a massless photon) and all fermions. 

In the Minimal Standard Model (MSM), the Higgs sector is the simplest possible: it contains two complex Higgs fields, which after giving masses to $W^+,W^-, Z^o$, leave a {\it single neutral scalar Higgs particle which should be discovered}. Matter is built of fermions - quarks and leptons, there are three families of each, with the corresponding antiparticles; quarks come in 3 colors; leptons are color singlets, they don't interact strongly. Bosons are carriers of interactions: 8 massless gluons, 3 heavy weak bosons ($W,Z$) and 1 massless photon. A massive neutral scalar Higgs field permeates the Universe and is (in some way) responsible for the masses of other particles (for each fermion, the same constant describes the fermion mass and its interactions with the Higgs field). Higgs scalar particle is the only particle not yet observed in the MSM. There are 26 parameters in MSM, not predicted by the theory: masses of quarks and leptons; coupling constants of $SU(3),SU(2)$ and $U(1)$; Higgs mass and vacuum expectation value; Cabibbo-Kobayashi-Maskawa quark mixing angles and a complex phase;   Maki-Makagawa-Sakata lepton mixing angles and a complex phase; and a QCD phase $\Theta$. All must be measured. Despite the MSM's remarkable success in describing all existing data so far, a long list of unanswered questions makes it obvious that MSM is but a low-energy approximation to a more complete theory. 
Why are there so many free parameters (all masses, all mixing angles and CP-violating phases)? Why are there equal number of quarks and leptons - is there an additional symmetry? Why do quarks and leptons exist in three pairs (generations)? Why is CP not an exact symmetry (or why are the laws of physics not symmetrical between matter and antimatter) - this is most likely related to the question why our Universe is matter-dominated. Are quarks and leptons elementary or do they have structure at a scale smaller than we can probe ($< 10^{-18}$ m)? The muon and electron look identical, except for their masses, could muon be an "excitation" of what constitutes what we currently think of as a "pointlike" electron? Are the neutrinos Dirac or Majorana particles? Why are their masses so small? Are protons stable? In QCD, the confinement of quarks and qluons was never proven - if we live in low temperature where confinement works - is there a phase transition at higher temperatures where quarks become free? What is the nature of the spontaneous symmetry breaking mechanism of electro-weak theory? Do strong and electroweak interactions become one at very high energies? And last, but not least: how to include gravity? There are many theories "beyond the Standard Model": Supersymmetry, Technicolor, Grand Unified Theories based on larger symmetry group, e.g. $SU(5),SO(10), E_8$, Monster group; string theory, superstring theory, branes, M-theory, extensions of Kaluza-Klein theory - new experimental data are needed badly to test any of  those new ideas. There are three ways to obtain the relevant experimental data: 
\begin{itemize}
\item  high precision low energy experiments, in which comparisons are made with precision higher order Standard Model calculations
\item discrepancies between cosmic ray and astrophysical measurements and their interpretation in the language of  Standard Model
\item by directly colliding particles accelerated to the highest possible energies in particle physics laboratories
\end{itemize}
\section*{Tevatron at Fermi National Accelerator Laboratory}

Accelerators are the microscopes of particle physics. In electron-positron colliders the colliding particles have no structure, the kinematics of the collision is known completely and all of the collision energy is transformed into the produced particles. Ideally, such machines would be the preferred tool to conduct high energy physics experiments. Unfortunately, there are technical problems. Electrons are difficult to accelerate, they undergo large energy losses when accelerated because of the small electron mass, which leads to either very long linear colliders (SLAC), or very large radius circular machines (LEP). Hadron circular colliders are much easier to build; however, they are "messy". The colliding particles are {\it not} elementary, they can be thought of as bags filled with quarks and gluons. The interaction takes place between the constituents of the colliding hadrons, each with an unknown fraction of the longitudinal component of the hadron momentum. Not all of the proton-proton or proton-antiproton energy is available for producing new particles. Tevatron at Fermilab and LHC at CERN are examples of such accelerators. The important parameters characterizing any accelerator is the beam type, beam energy (or, rather, energy available in the collision) and luminosity (related to beam intensity). Tevatron is a superconducting proton-antiproton synchrotron accelerator. It has 774 superconducting dipole magnets with 4.2 T magnetic field and 240 superconducting quadrupole magnets arranged in a 6.28 km ring. The first accelerated beam at the energy of 512 GeV was produced in 1983, and there were many upgrades to the machine since then. The collider data taking periods and their characteristics are listed in Table~\ref{table::runs}.

\begin{table}[ht]
\caption{Tevatron running periods}
\centering
\begin{tabular}{l l l c}
\hline
		&years		&delivered luminosity	& $\sqrt s $ (cms energy) \\
Run 0	&1988-1989	&~~10/pb				&1800 GeV \\
Run I	&1992-1996	&120/pb				&1800 GeV \\
Run II	&2001-2011	&~~10/fb				&1960 GeV
\end{tabular}
\label{table::runs}
\end{table}

\begin{figure}[h]
\hspace{0.0cm}
\includegraphics[width=0.95\textwidth]{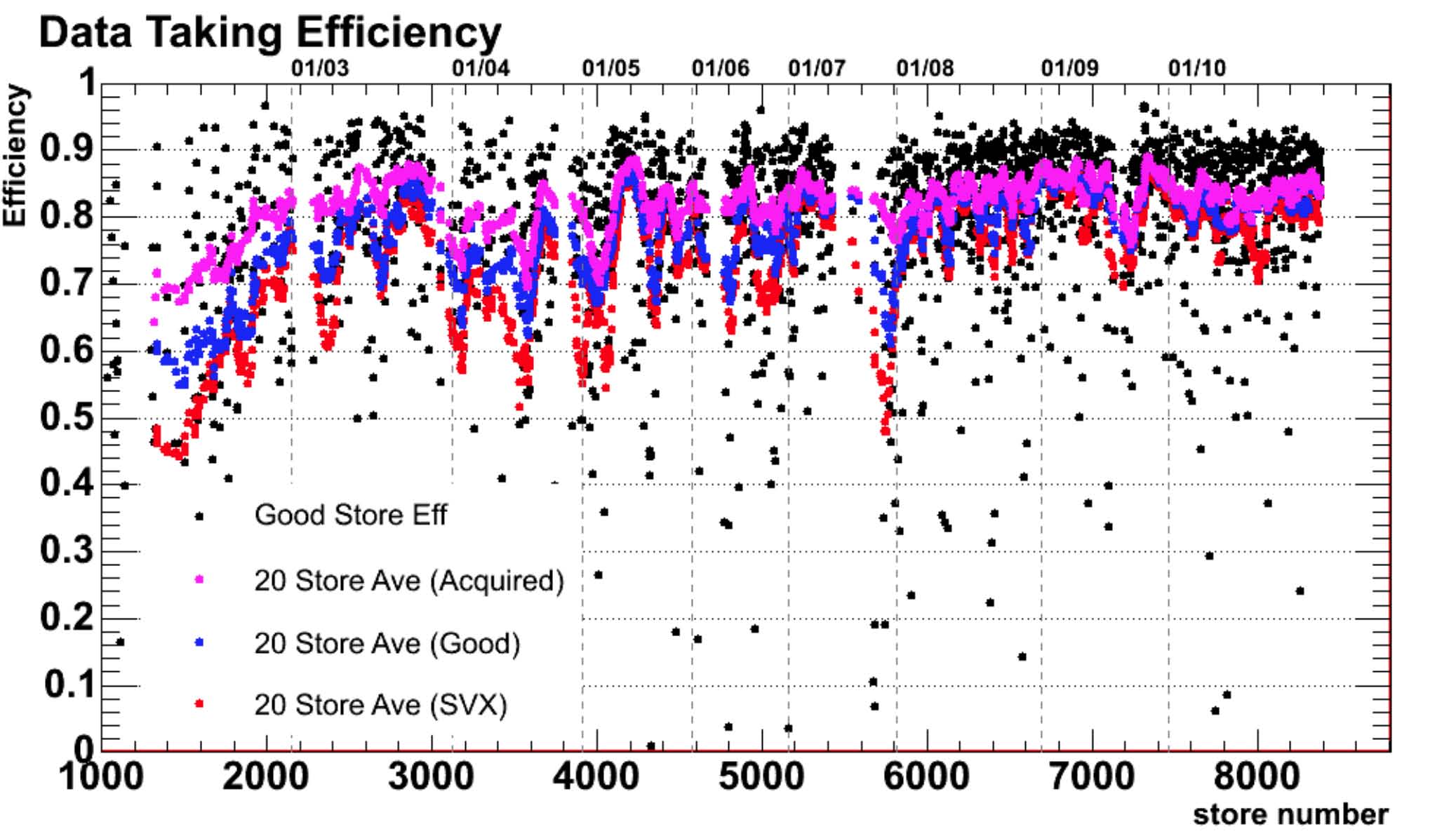}
\includegraphics[width=1.03\textwidth]{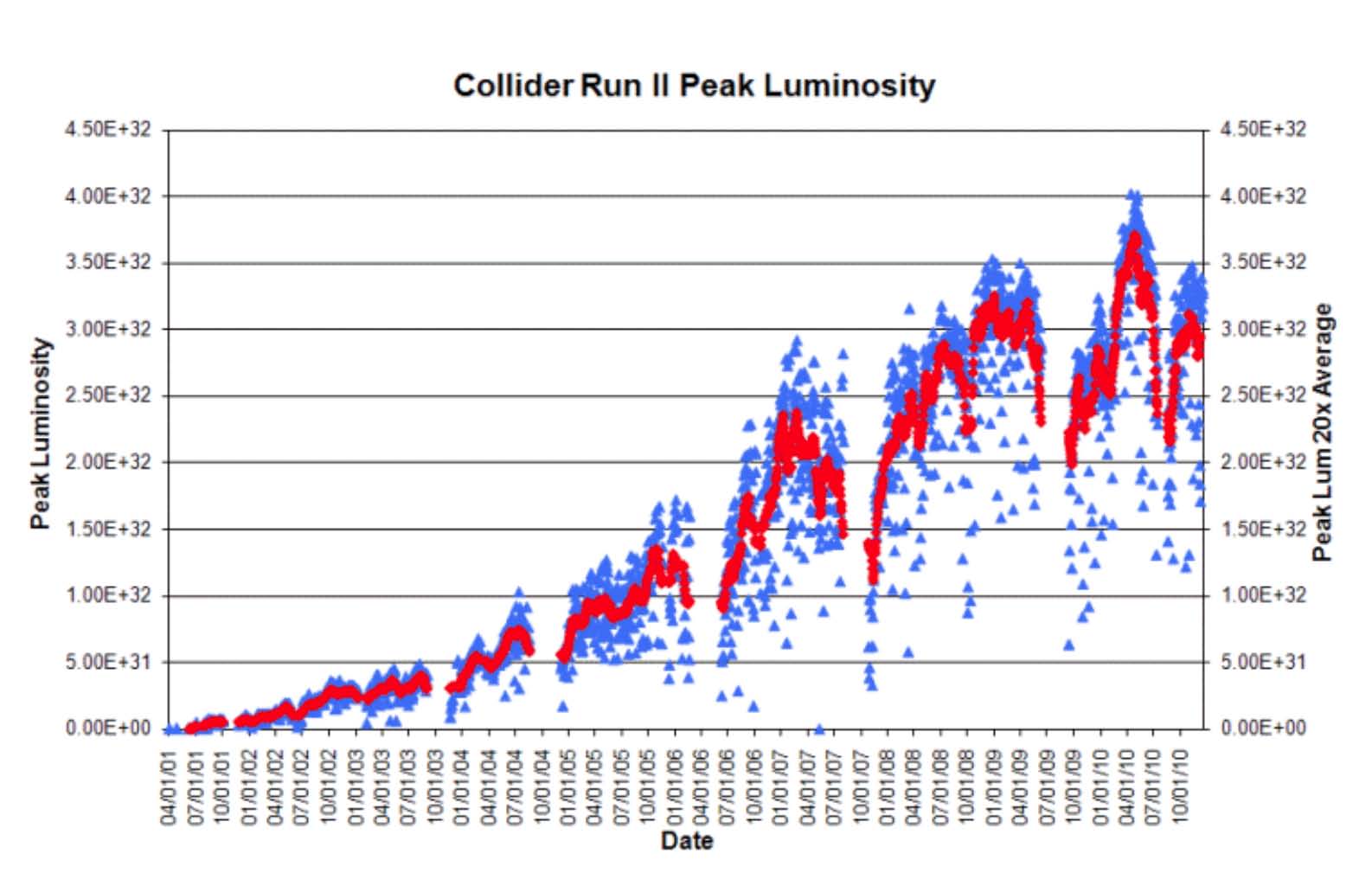}
\caption{\label{fig:data_eff} \small Tevatron performance in Run II as a function of time or store number (each time the ring is filled with colliding protons and antiprotons is called a store). SVX efficiency requires that the silicon vertex detector is operational, GOOD means that all other detectors but SVX are running well. In the plots, points and triangles indicate the peak luminosity at the beginning of each store, the lines show the running averages of luminosity, averaged over 20 stores.}
\end{figure}

\begin{figure}[h]
\vspace{-0.5cm}
\hspace{-0.7cm}
\includegraphics[width=0.63\textwidth]{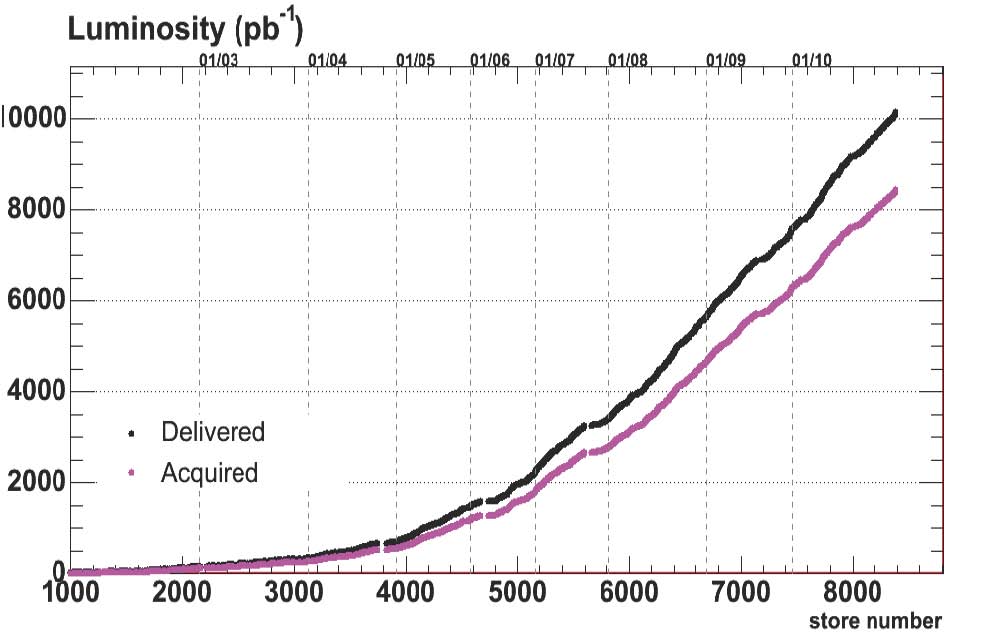}
\includegraphics[width=0.41\textwidth]{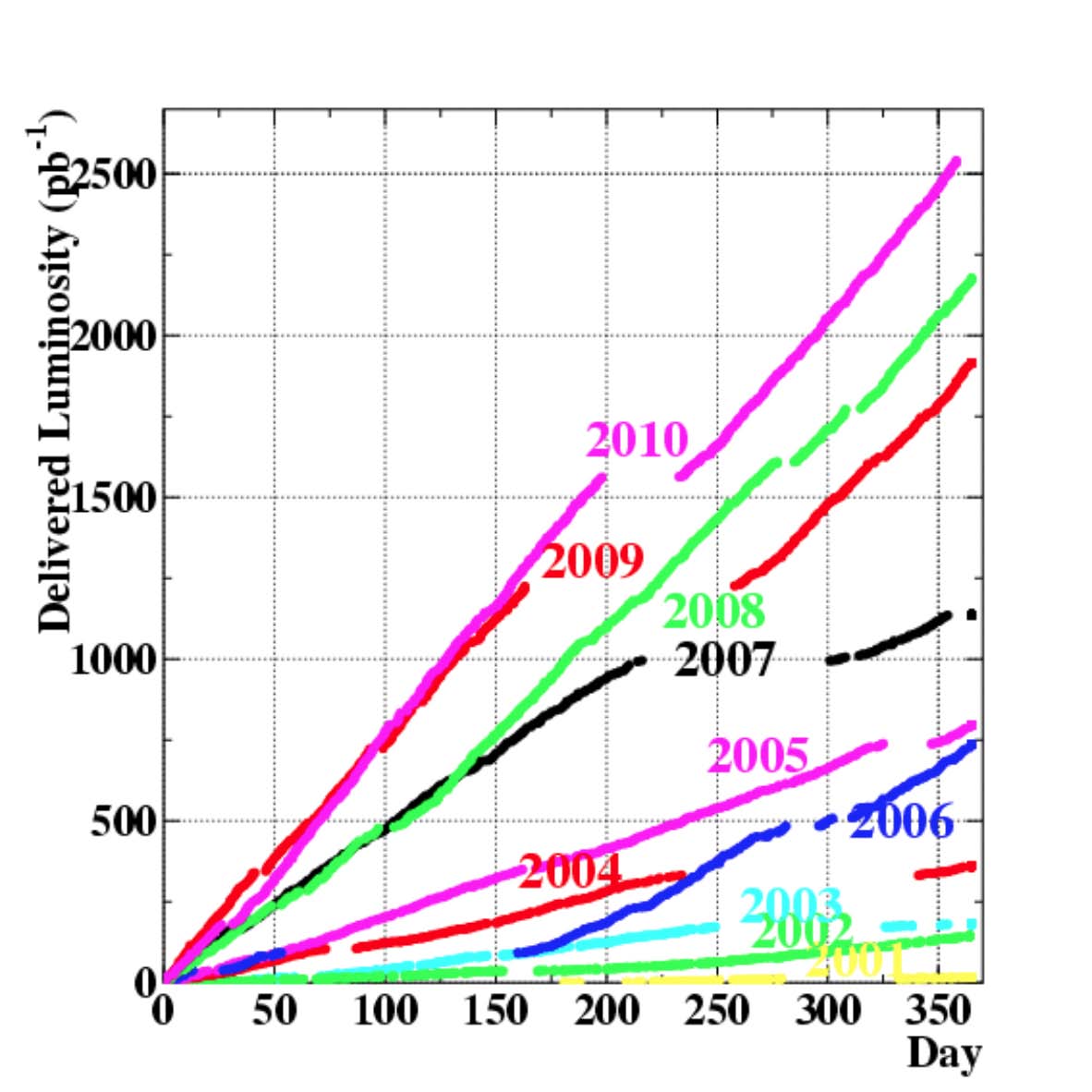}
\caption{\label{fig:lum} \small Left: Integrated luminosity in Tevatron Run II as a function of time. Right: Delivered luminosity in Run II year-by-year.}
\end{figure}

The most significant upgrade among those performed for Run II was construction of the new Main Injector. The result was an increase of the centre-of-momentum (cms) energy from 1800 to 1960 GeV, increase of luminosity by orders of magnitude, and a reduction of the beam crossing time from 3.5 $\mu s$ (in Run I) to 396 ns, which made possible significant increases in instantaneous luminosity while keeping the number of multiple interactions low. (Even this seemingly modest change of energy leads, for example, to an increase of the $t{\bar t}$ cross section by 35\%). 
\noindent
Figures~\ref{fig:data_eff},\ref{fig:lum} show the data taking efficiency (for CDF), peak luminosity in Tevatron Run II, integrated luminosity as a function of time, and the Run-II delivered luminosity year-by-year. It is impressive to see the peak luminosity exceeding $3 \times 10^{32}$ cm$^{-2}$s$^{-1}$ in the last 2-3 years of the machine operation (the 1982 design luminosity was $10^{30}$ cm$^{-2}$s$^{-1}$). However, one should notice that it took more than 3 years to learn how to operate the Tevatron after restarting with its Run II upgrades, the fact that should be remembered when judging the progress of LHC startup. It is also clear from the plots that in the last few years Tevatron has reached its performance limits. After 10 years of running, in the week of December 13, 2010, the integrated delivered luminosity exceeded 10/fb (8/fb acquired) per experiment. At the same time, each additional year of running can deliver no more than $\sim$2-2.5/fb per experiment.


\section*{The detectors}

Two large, multipurpose detectors, CDF and D0~\cite{det}, have been designed and built to collect and analyze the data from the proton-antiproton collisions at the Tevatron. Both detectors underwent significant upgrades for Run II. D0 has added a silicon vertex detector (SVX) to allow better b-quark tagging, and also installed a solenoid to allow track momentum reconstruction. A high resolution liquid argon and uranium calorimeter was a strength of D0 already in Run I. D0 routinely reverses the orientation of its magnetic field, which allows cancellation of many possible detector effects, an important feature for any charge asymmetry measurement. CDF has upgraded its calorimeter in the "plug" region, which significantly improved the energy resolution in the pseudorapidity range of $1.1 < \vert \eta \vert < 3.5$. CDF also installed a new (longer) SVX which doubled the Run-I b-quark tagging efficiency. Both D0 and CDF are large collaborations of $\sim 500$ physicists each. 

\section*{Tevatron physics program}

The most important physics goals of the Tevatron program were: 
\medskip

\noindent
{\bf ~i) searches for physics "beyond the Standard Model"}

\noindent
{\bf ii) precision measurements and tests of the Standard Model: }
\begin{itemize}
\item
QCD studies
\item W mass measurement
\item b-quark physics (lifetimes, spectroscopy, CP violation studies..)
\item $WW,WZ,ZZ,W\gamma,Z\gamma, \gamma\gamma$ physics
\item top quark physics (top quark has been discovered at Tevatron in 1993-1994
\item Higgs searches (MSM, Minimal Supersymmetric Standard Model)
\end{itemize}

\bigskip

\begin{figure}[h]
\vspace{-3.4cm}
\hspace{-1.5cm}
\includegraphics[width=0.5\textwidth]{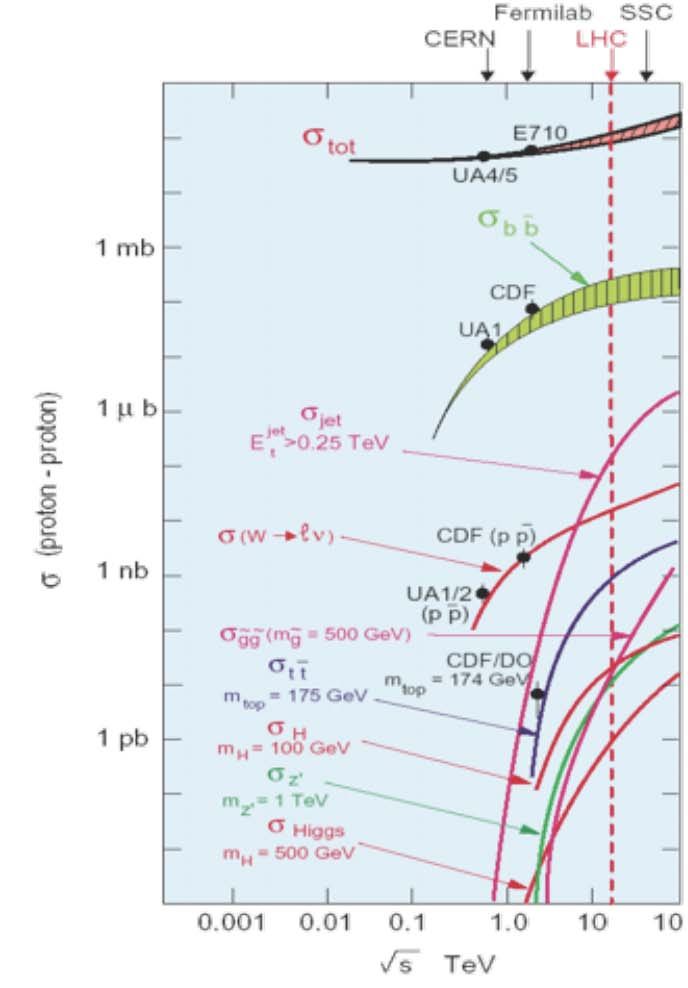}
\includegraphics[width=0.7\textwidth]{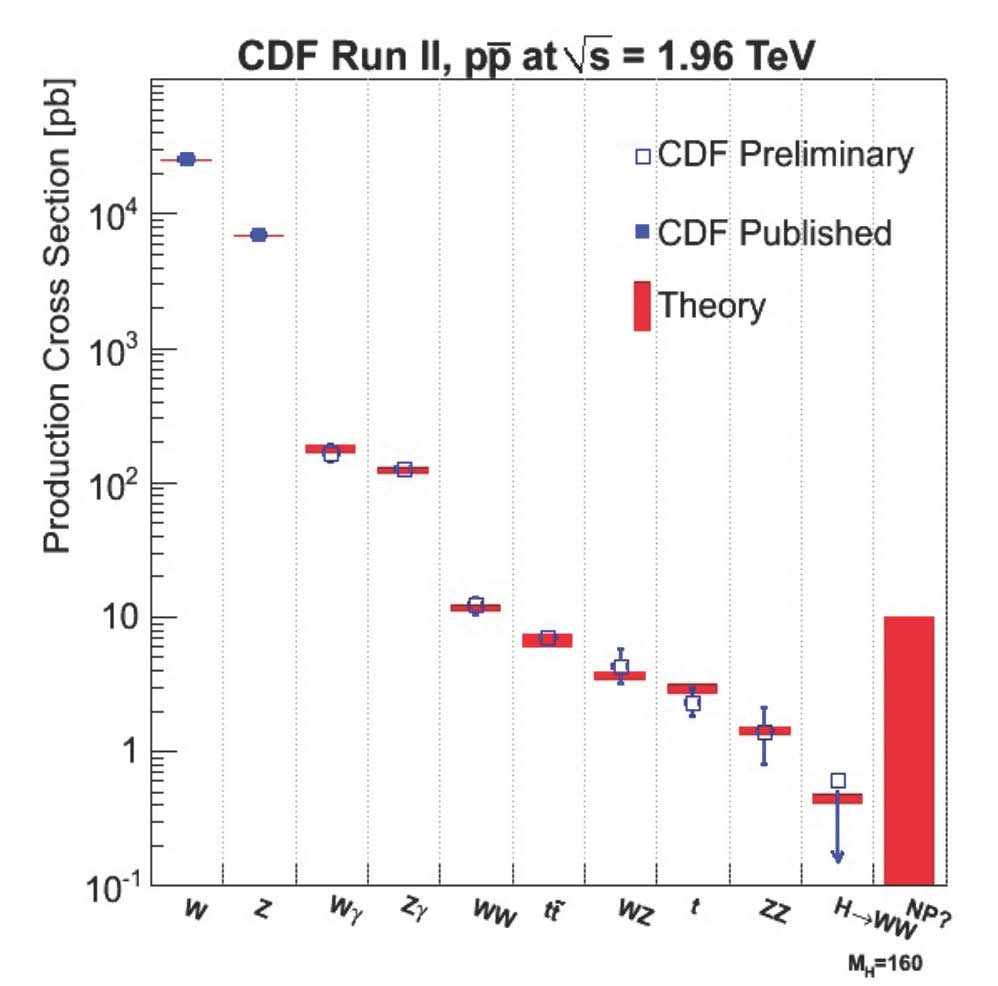}
\caption{\label{fig:sigmas} \small Cross sections for selected processes at SpS, Tevatron and LHC}
\end{figure}

A comparison of the physics reach of CERN SpS, Tevatron and LHC is presented in Figure~\ref{fig:sigmas}, in which the cross sections for most of the processes listed above are shown as a function of cms energy. Also shown (on the right) is a more detailed summary of a number of selected Standard Model production cross sections at the Tevatron Run II,  together with the measured values from CDF, when available. At present, with $\sim 10$/fb of data available per experiment, Tevatron can probe processes with cross sections of the order of a picobarn. CDF and D0 can thus perform studies of QCD and the electroweak part of SM, multi-boson couplings, single top production and, if not to observe, at least to place meaningful limits on MSM Higgs production. Tevatron is also a b-quark factory, allowing important studies of CP -violation in $B^0$ system, and spectroscopy of $B$ mesons and baryons. Because of space limitations, I will show only a selection of D0 and CDF results (some of the results became already obsolete with $\sim$ 35-45/pb of data collected at the LHC in 2010). Also, when discussing physics analyses and results I shall put emphasis on methodology at the expense of many analysis details, all of which can be found on the publicly accessible web-sites~\cite{D0results}\cite{CDFresults}.

\subsection*{Searches for new particles}

Here, the much higher energy available at the LHC, $\sqrt s=$7 TeV as opposed to $\sqrt s =$1.96 TeV at Tevatron, combined with the parton distribution functions (PDF), provides a decisive advantage for LHC experiments. Most of the current Tevatron limits on existence of new heavy particles (based on $\sim$1-3/fb of data) have already been made obsolete by LHC limits (based on only $\sim$3-30/pb of data).  For example, for 400 GeV gluinos (if existed), the predicted cross section would result in a smaller number of produced events for 4/fb at Tevatron than for 3/pb at LHC. CDF limits for the mass of excited quarks (looking for a bump in dijet mass distribution) is $M(q^*)>870$ GeV/c$^2$; the correponding LHC limits based on $\sim$3/pb of data are $M(q^*)>1.5$ TeV/c$^2$ (ATLAS) and 
$M(q^*)>1.58$ TeV/c$^2$ (CMS). Most limits on Supersymmetry set by searches at Tevatron are already obsolete with the 2010 data from LHC ($\sim$40/pb).

\subsection*{QCD studies}
\begin{figure}[h]
\vspace{0.0cm}
\hspace{0.7cm}
\includegraphics[width=0.7\textwidth]{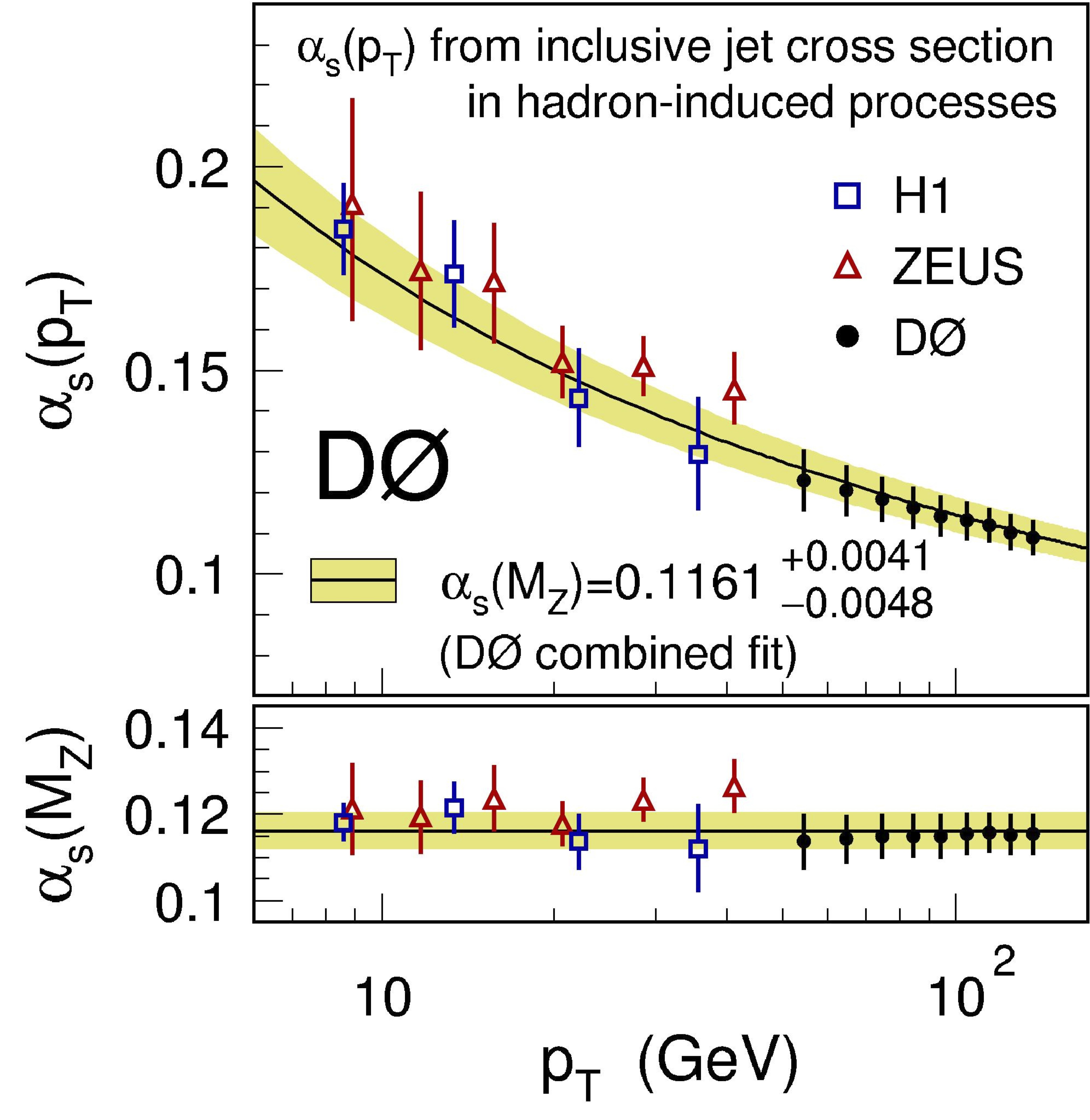}
\caption{\label{fig:alphas} \small D0 measurement of $\alpha_s(M_Z)$ based on 0.7/fb of data.}
\end{figure}

A measurement of the strong coupling constant, $\alpha_s$, has been made by both D0 and CDF. The D0 measurement\cite{QCD} is shown in Figure~\ref{fig:alphas}. The analysis uses the transverse momentum, $p_T$, dependence of the jet cross section. The analysis is based on a fit to the inclusive jet cross section in the range ($50 < p_T <145)$ GeV/c, after excluding high $p_T$ points to mimimize the PDF uncertainty correlations. The NLO and 2-loop threshold corrections, and MSTW 2008 NNLO parametrization of PDF's have been used. The result is: 
$\alpha_s(M_Z)=0.1161^{+0.0041}_{-0.0048}$
to be compared with an earlier CDF result\cite{CDF_QCD}:  $\alpha_s(M_Z)=0.1178^{+0.0081}_{-0.0095}.$

\subsection*{W boson mass}
The measurement of of W mass is based on a fit to the transverse mass of a lepton and transverse missing energy, \MET. Although in principle very simple, this analysis requires a very good understanding of modeling of the recoil system, and understanding of  background and numerous systematics effects. Not surprisingly, the systematic errors are comparable with the statistical uncertainties. The current Tevatron average value\cite{Wcomb}  (from 2009, based on analyses of $\sim$1/fb of data), is:
\begin{eqnarray*}
M_W=80.420\pm0.031~ GeV/c^2
\end{eqnarray*}

\begin{figure}[h]
\vspace{-0.1cm}
\hspace{-1.2cm}
\includegraphics[width=0.6\textwidth]{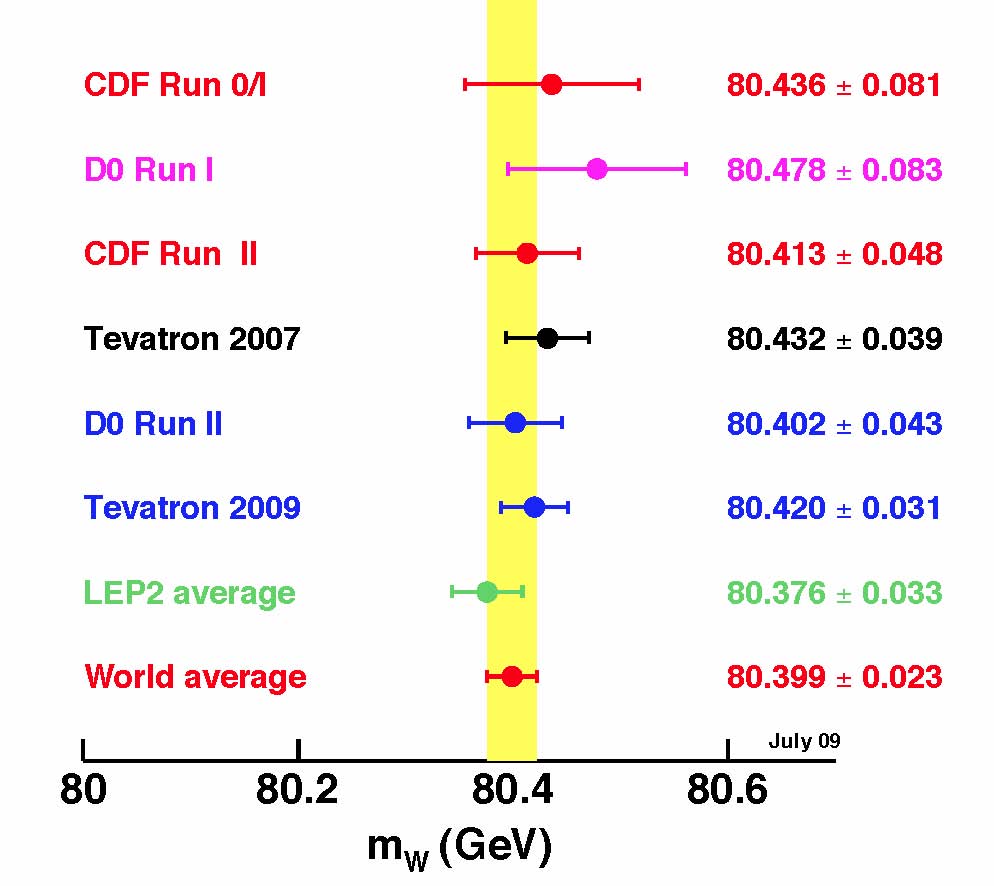}
\includegraphics[width=0.6\textwidth]{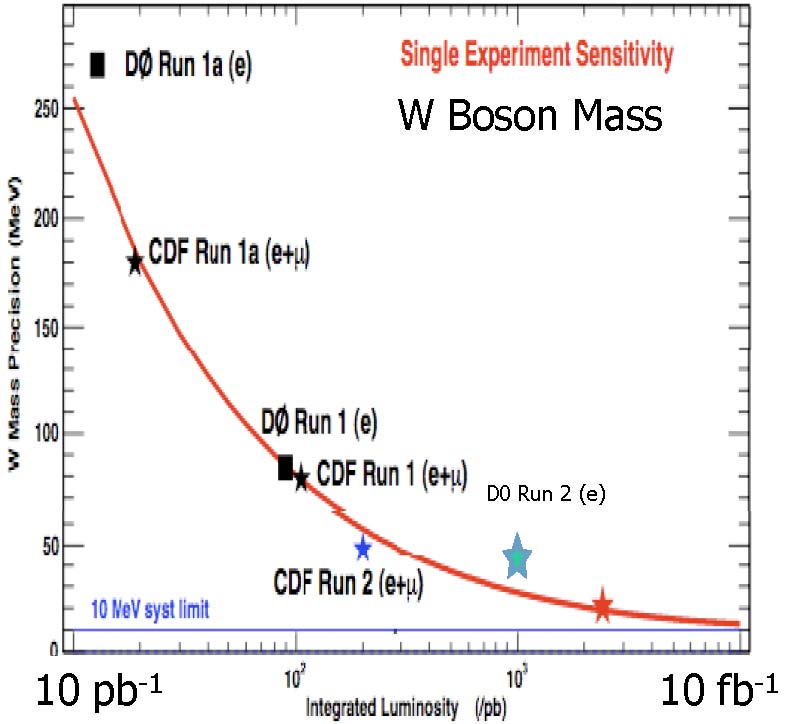}
\caption{\label{fig:WM} \small W mass measurements, most recent CDF Run II result is based on 0.2/fb, while D0 Run II result is based on 1/fb. On the right side, predicted errors for W mass measurement at Tevatron.}
\end{figure}

\noindent
A summary of $M_W$ measurements available from all experiments is shown in Figure~\ref{fig:WM}. Tevatron plans to reach $\Delta M_W\sim 20$ MeV/c$^2$, after combining D0 and CDF results based on their full datasets, when they become available. CDF results based on $\sim$2.3-2.4/fb  demonstrate that the statistical errors of the order of 15 MeV/c$^2$ are already achievable. The current projections for the W mass measurement errors for a single experiment are shown in the same Figure, on the right. The W mass measurement could be one of the most important physics results from the Tevatron, and it may be the longest lasting legacy of the Tevatron program. It will be quite difficult for the LHC experiments to reach the W mass precision comparable to that of Tevatron. The reason is that the LHC is a proton-proton machine while the Tevatron is a proton-antiproton accelerator. Our understanding of PDF's is much better for valence quarks than that for the parton "sea" in a proton, and this situation may not change for quite a number of years.

\subsection*{CP violation in $B_s \to J/\psi \phi$}

\begin{figure}[h]
\vspace{-0.2cm}
\hspace{-2.0cm}
\includegraphics[width=0.65\textwidth]{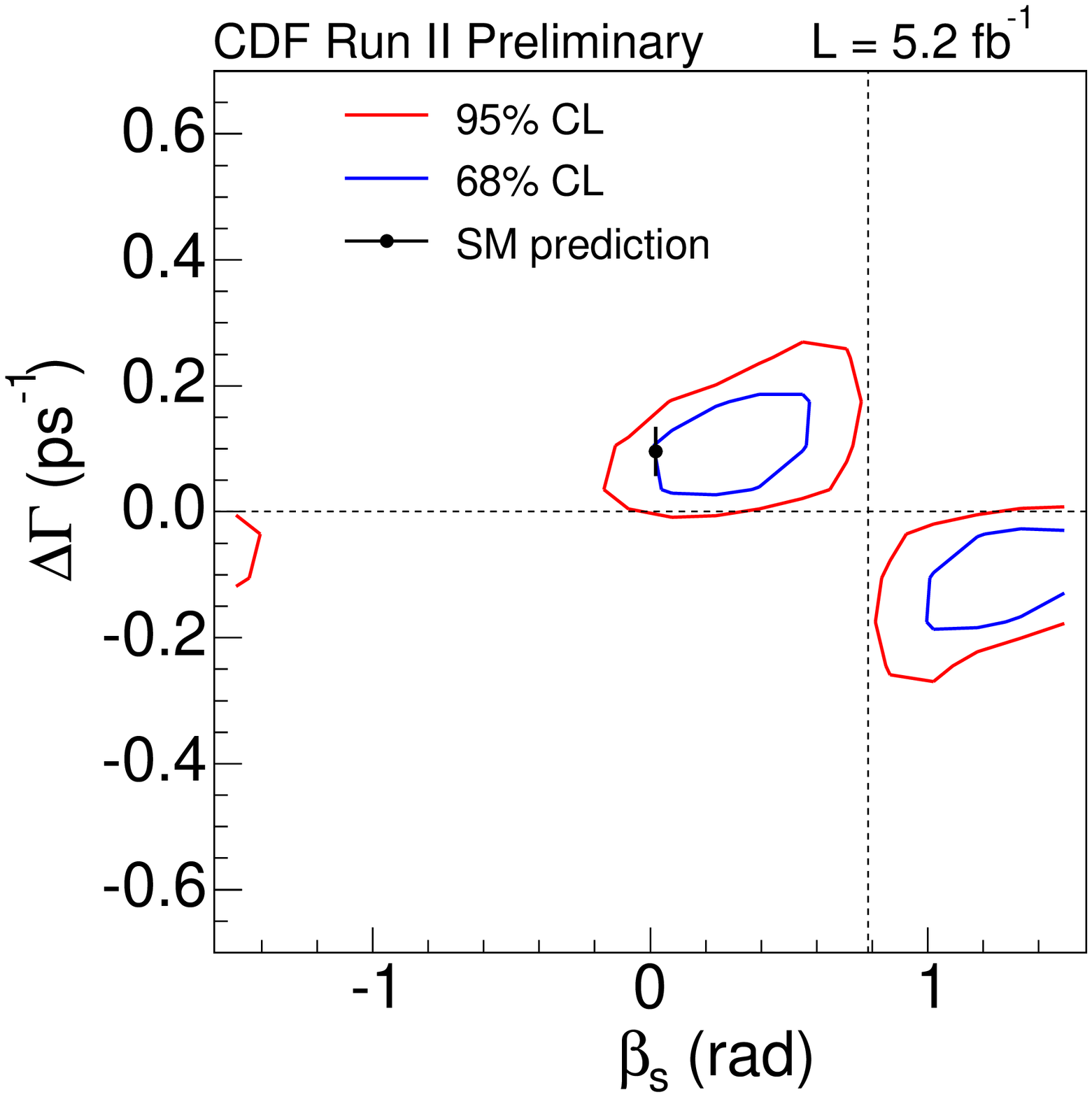}
\includegraphics[width=0.68\textwidth]{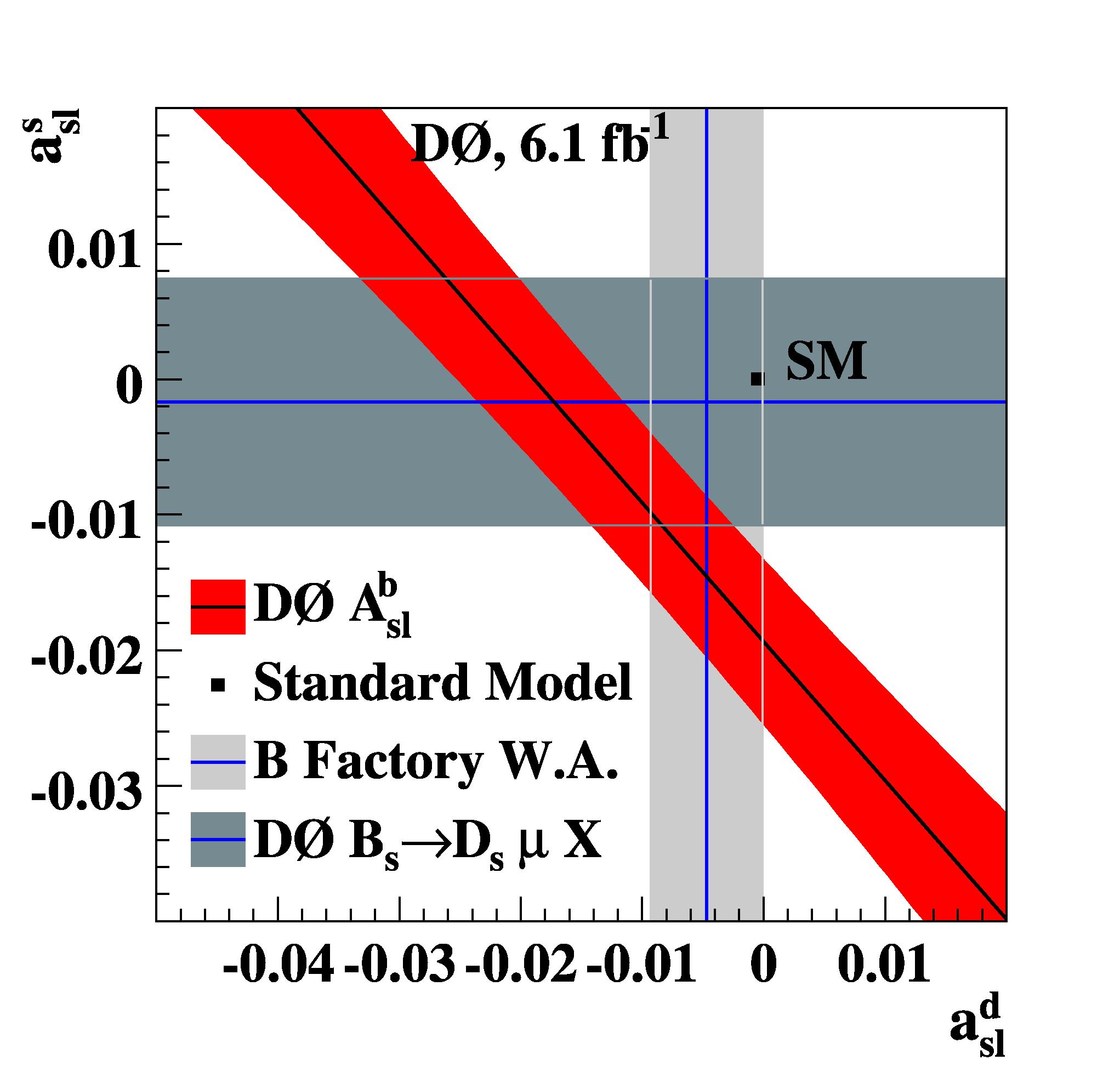}
\caption{\label{fig:CDF_CP} \small Left: CDF $68\%$ and $95\%$ confidence regions in the $\beta_s-\Delta\Gamma$ plane. Assuming the standard model predictions of $\beta_s$ and $\Delta \Gamma$ , the probability of a deviation as large as observed data in the data is $44\%$, corresponding to $0.8\sigma$. Right: Comparison of D0 $A^b_{sl}$ in data with the SM
prediction. The bands represent the $1\sigma$ uncertainties.}
\end{figure}

The studies of discrete symmetries - charge conjugation (C),
parity (P) and time reversal (T), have provided important information about the structure of weak interactions. The violation of CP symmetry has only been observed in the weak interactions. CP violation has been well established in the $K^0$ and $B_d^0$
systems, but not in the $B_s^0$
system where the effects of CP violation
are expected to be small in SM. The observed
magnitude of CP violation in the $K^0$ and $B_d^0$
systems is not
sufficient to explain the observed matter-antimatter asymmetry in the Universe, suggesting the
presence of additional sources of CP violation beyond the
SM. At the Tevatron, the initial $p{\bar p}$ state is CP invariant. In addition, $b$ quarks are produced
mainly in $b{\bar b}$ pairs. This fact makes it possible to use flavor tagging to distinguish whether a $B_s^0$ or  
${\bar B_s^0}$ were produced. CDF uses flavour tagging in order to maximize the sensitivity in its measurement\cite{CP} of the CP-violating phase $\beta_s^{J/\psi \phi}$. One can determine the flavor of the $B_s^0$ at production stage by looking at tracks associated with the hadronization process of the $b/{\bar b}$ quark that produced the observed $B_s^0$/${\bar B_s^0}$ (same-side tag or SST), or by looking at the decay products of $B$ hadrons produced by the other $b/{\bar b}$ quark in the event (opposite-side tag or OST). Different tagging algorithms are used to tag the flavor of the opposite-side $b$ quark, depending on whether the opposite side decay products are electrons (soft electron tagger), muons (soft muon tagger), or jets (jet charge tagger). CDF then uses a Neural Network (NN) combination of the OST for more optimal tagging power. The efficiency of OST is $94\pm1\%$ and that of SST is $52\pm1\%$. The new measurement,  $\Delta \Gamma=0.075\pm0.035(stat)\pm0.01(syst)$ ps$^{-1}$, is based on 5.2/fb of data. It agrees with SM prediction better than a similar analysis based on a smaller dataset and is shown on the left plot in Figure~\ref{fig:CDF_CP}.

 As mentioned earlier, D0 experiment
is particularly well suited to study the small effects of CP violation because the periodic reversal of the
D0 solenoid and toroid magnetic field polarities results in a
cancellation of most detector-related charge asymmetries.
D0 measures CP violation in $B_s$ and $B_d$ mixing using  the charge asymmetry for like-sign muon pairs in semileptonic $B$ decays: $A^b_{sl}\equiv(N_b^{++}-N_b^{--})/(N_b^{++}+N_b^{--})$, where $N_b^{++},N_b^{--}$ are the numbers of events with two $b$ hadrons decaying semileptonically, leading to two same charge muons in the final state. One muon comes from direct semileptonic decay $b\to\mu^- X$, the second muon is assumed to come from direct semileptonic decay after a neutral $B^0$ meson mixing into ${\bar B^0}$. Based on the dataset corresponding to an integrated luminosity of 6.1/fb, D0 finds evidence for anomalous like-sign charge asymmetry\cite{D0asym}. The measured value of 
\begin{eqnarray*}
A^b_{sl}=(-0.957\pm0.251(stat)\pm0.146(syst))\times10^{-2}
\end{eqnarray*}
\noindent
is $3.2\sigma$ from SM prediction $A^{theory}_{sl}=(-2.3\pm0.6)\times10^{-4}$, and is shown on the right plot in Figure~\ref{fig:CDF_CP}.
This measurement could be the first evidence for "beyond the Standard Model" CP violation.

\subsection*{Spectroscopy}

\begin{figure}[h]
\vspace{-0.5cm}
\hspace{-0.9cm}
\includegraphics[width=0.61\textwidth]{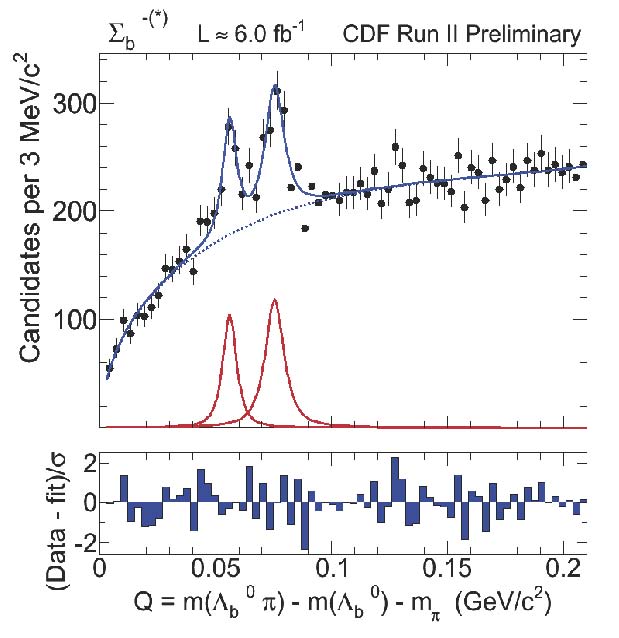}
\includegraphics[width=0.61\textwidth]{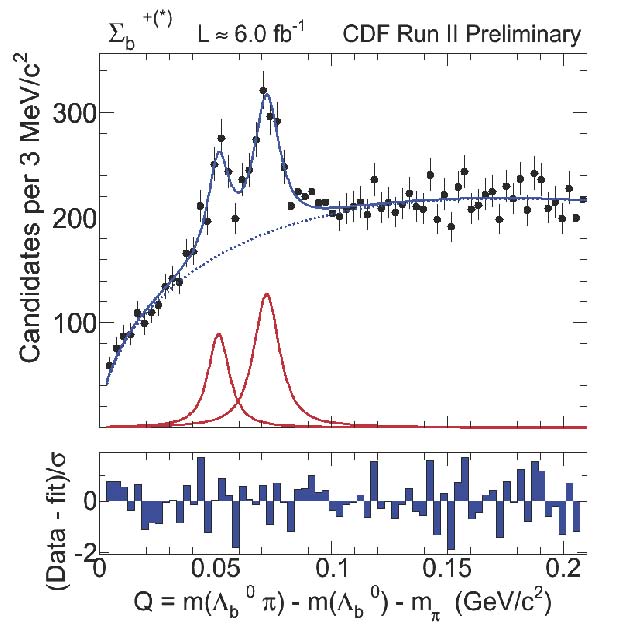}
\includegraphics[width=0.5\textwidth]{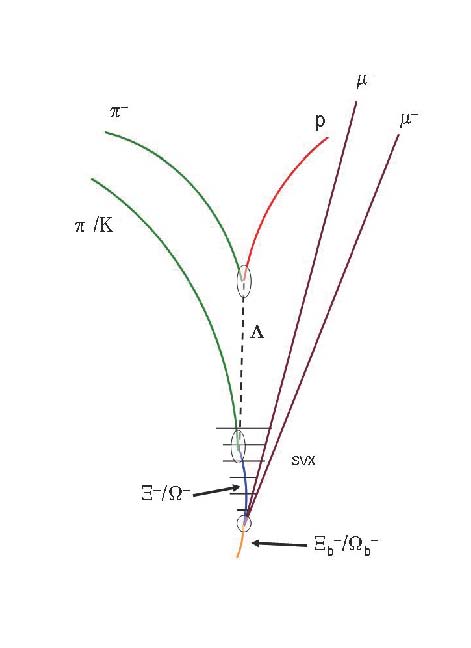}
\hspace{0.5cm}
\includegraphics[width=0.5\textwidth]{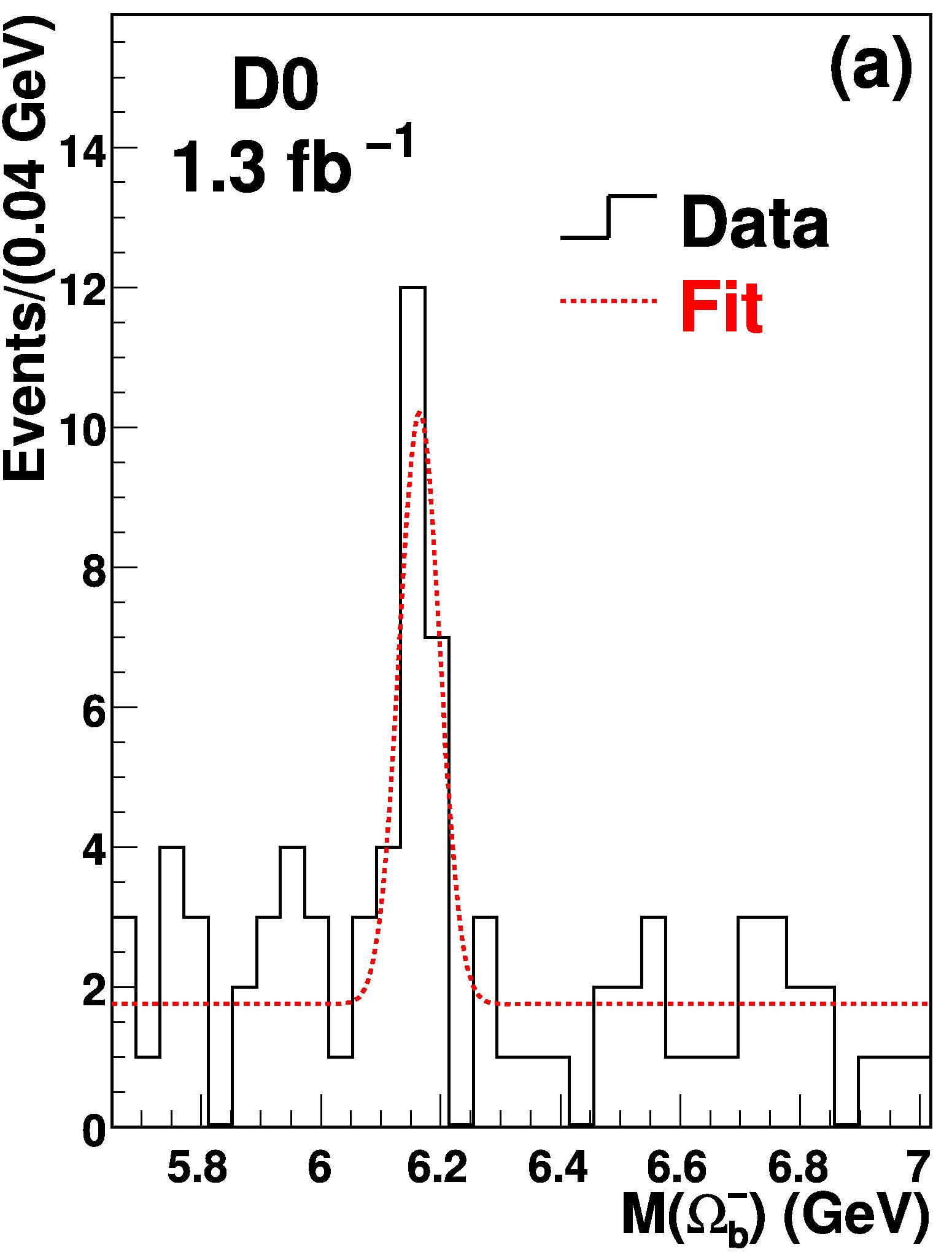}

\caption{\label{fig:BB} \small Mass distributions showing $\Sigma_b$,$\Sigma_b^*$, and $\Omega_b$, together with a diagram of a cascade decay identified in the analysis.}
\end{figure}

\begin{figure}[h]
\vspace{-0.5cm}
\hspace{-1.1cm}
\includegraphics[width=0.6\textwidth]{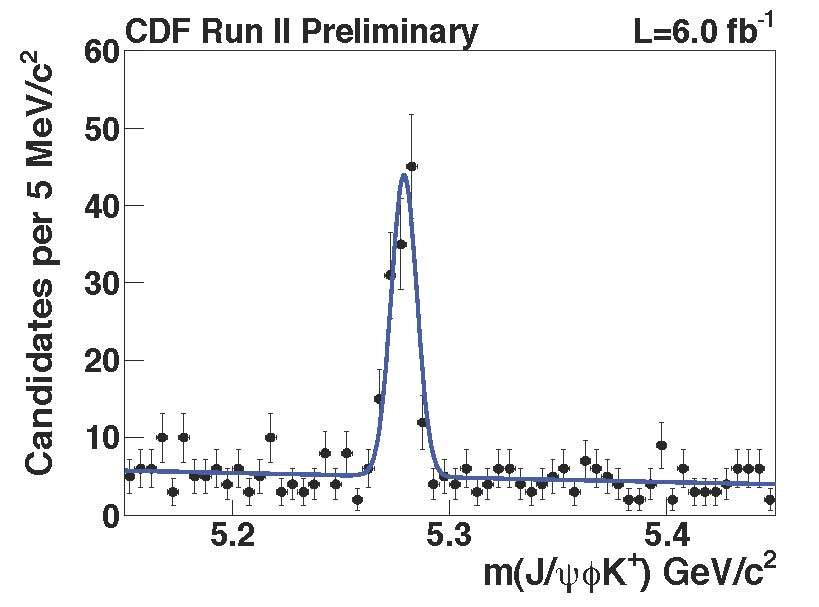}
\includegraphics[width=0.6\textwidth]{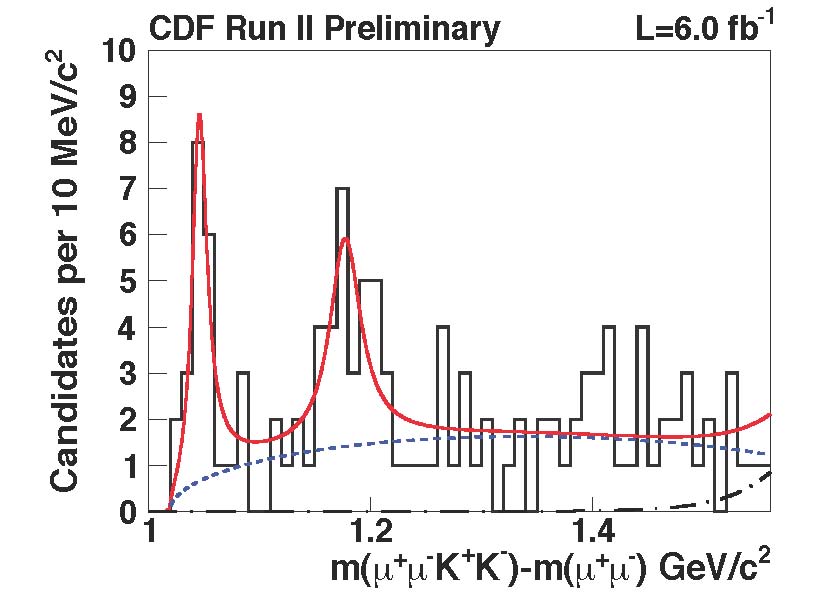}
\caption{\label{fig:CDFY} \small Mass distributions of  $J/\psi \phi K$ system, and mass difference of $\mu^+\mu^-KK$ and $\mu^+\mu^-$ systems.}
\end{figure}

Large production cross sections for mesons and baryons containing b-quark allowed observations of $\Sigma_b$'s (2006)~\cite{Sigma}, $\Xi_b$ (2007) and $\Omega_b$ (2008)~\cite{Omega}. The search strategies rely on ability to suppress backgrounds by selecting decay modes involving $J/\psi$ and identifying the secondary vertices using SVX detectors. The results are shown in Figure~\ref{fig:BB}, and the characteristics of the b-quark baryons, as measured at Tevatron\cite{spectr}, are listed below:
\begin{eqnarray*}
m(\Sigma_B^+)&=&5811.2^{+0.9}_{-0.8}(stat.)\pm1.7(syst.)~MeV/c^2 \\
m(\Sigma_B^-)&=&5815.5^{+0.6}_{-0.5}(stat.)\pm1.7(syst.)~MeV/c^2Ê\\
m(\Sigma_B^{*+})&=&5832.0\pm0.7 (stat.)\pm1.8(syst.)~MeV/c^2\\
m(\Sigma_B^{*-})&=&5835.0\pm0.6(stat.)\pm1.8(syst.)~MeV/c^2\\
m(\Xi_b)&=& 5790.9\pm2.6(stat.)\pm0.8(syst.)~MeV/c^2\\
\tau(\Xi_b)&=&1.56^{+0.27}_{-0.25}(stat.)\pm0.02(syst.)~ps\\
m(\Omega_b^+)&=&6054.\pm6.8(stat.)\pm0.9(syst.)~MeV/c^2\\
\tau(\Omega_b^+)&=&1.13^{+0.53}_{-0.40}(stat.)\pm0.02(syst.)~ps 
\end{eqnarray*}
CDF has also performed searches for exotic QCD mesons, either multi-quark meson molecules ($c{\bar c}u{\bar u}$) or hybrid mesons - quark-antiquark-gluon ($c{\bar c}g$), looking for structures in $B\to J/\psi \phi$ mass spectrum above the threshold at $4.116~GeV/c^2$. Experimentally, it is easy to search for such structures in a clean $B\to J/\psi \phi K$ channel taking advantage of the finite $B$ lifetime and narrow $B$ mass window. The results are shown in Figure~\ref{fig:CDFY} and the characteristics of the Y(4140) and Y(4270) states, observed in the decay chain 
\begin{eqnarray*}
 B^+\to Y(4140)K^+; Y(4140) \to J/\psi \phi; J\psi \to \mu^+\mu^-; \phi \to K^+K^-
\end{eqnarray*}

\noindent
are listed below\cite{Y}:
\begin{eqnarray*}
M_{Y(4140)}& =& 4143.4\pm3.0 (stat)\pm0.6 (syst)~MeV/c^2	\\
\Gamma_{Y(4140)} &=& 15.3\pm10.46.1(stat)\pm2.5 (syst)~MeV/c^2 \\
M_{Y(4270)}& =& 4274.4\pm 8.46.7 (stat)~MeV/c^2~~~(> 5 \sigma)\\
\Gamma_{Y(4270)} &=& 32.3\pm21.915.3  (stat)~MeV/c^2~~~ (  3.1 \sigma)
\end{eqnarray*}
\subsection*{Diboson production}
\begin{figure}[h]
\vspace{-0.9cm}
\hspace{-1.5cm}
\includegraphics[width=0.65\textwidth]{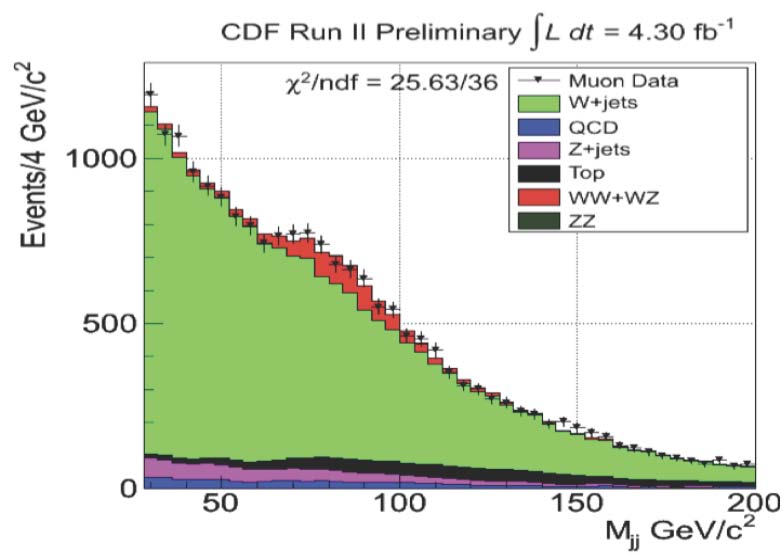}
\includegraphics[width=0.61\textwidth]{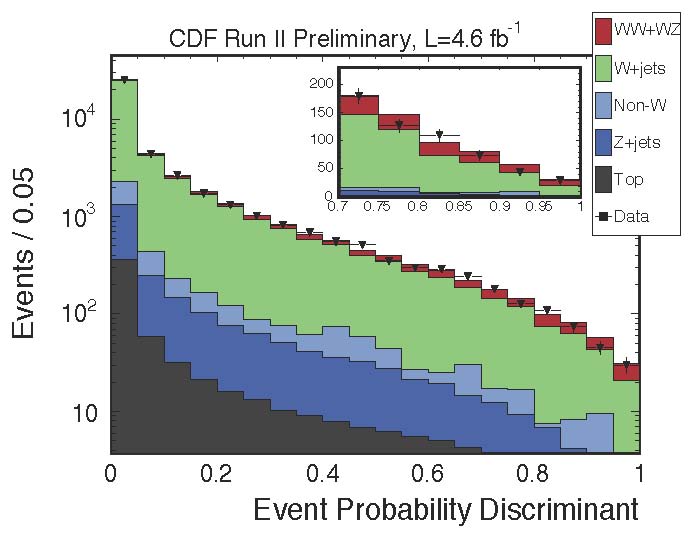}
\caption{\label{fig:WW} \small Left: Mass of two jets in events passing the $W$jets selection with an excess due to $W$ and $Z$ clearly visible. There is also a hint of another, smaller, excess of events  in the $140-150~GeV/c^2$ mass range. Right: Analogous measurement based on event discriminant, an output of a Neural Network (NN) based on a matrix element technique.}
\end{figure}

Searches for associated $WW,~WZ,~ZZ,~Z\gamma$ production are much more difficult because of their small cross sections, but very interesting - di-boson production probes boson couplings and is one of the best places to look for discrepancies between data and theory predictions which could provide signature of physics beyond SM. At the same time, those cross sections, while small, are larger than, for example, MSM Higgs production cross sections. Measuring di-boson production is thus necessary to gain confidence in complex analysis techniques and their application to measurements of very small signals with large backgrounds. Neural nets  and other multivariate techniques are routinely used in these analyses by both CDF and D0. In Figure~\ref{fig:WW} the mass of two jets in events with a $W$ is shown, an excess of events due to $WW$ and $WZ$ is observed {\it directly} in the $W$,$Z$ mass range. The other plot shows the event probability discriminant, based on matrix element (ME) technique. Here, one also identifies the $WW$ and $WZ$ contribution as an excess of events, except that here it is done {\it indirectly}. The CDF and D0 measurements of diboson cross sections\cite{WW}, all based on $\sim 4.5-6.4/pb$ of data, are:
\begin{eqnarray*}
\sigma(WW)& =& 12.1\pm0.9 (stat)\pm1.6 (syst)~ pb \\ 
\sigma(WZ)& =& 4.1\pm0.7~ pb\\
\sigma(Z\gamma)& = &4.6\pm0.2 (stat)\pm0.3 (syst)\pm0.3 (lum)~pb \\
\sigma(ZZ)& =&1.40\pm^{0.43}_{0.37} (stat)\pm 0.14 (syst)~ pb
\end{eqnarray*}
\subsection*{Top mass and top quark production cross section }
Search for top quark was one of the main physics goals of Tevatron collider program. Its existence was predicted within SM as a missing partner of b-quark in the weak isospin doublet in the third family of quarks. Top quark was discovered at Tevatron in 1993-1994~\cite{Top}. To measure the cross section one needs to:	i. search for events with top signature; ii. calculate expected SM background; iii. count events above backgrounds; iv. apply corrections for acceptance and reconstruction inefficiencies  and biases. However, some of the acceptance corrections are strongly varying functions of the top quark mass, $M_{top}$. The measured cross section thus depends on the adopted value of $M_{top}$, which has to be determined independently.

All mass measurement techniques assume that each selected event contains a pair of massive objects of the same mass (top and anti-top quarks) which subsequently decay as predicted in SM. 
It is assumed that the selected sample of events contains just the $t{\bar t}$ events and the SM background. This is the simplest and the most natural hypothesis since top quark is expected in SM.
The combinatorics, i.e. the problem that only one out of a large number of jets-lepton(s) combinations is correct, adds to the complexity of the problem.

In the lepton+jets and all-jets final states there is enough kinematical constraints to perform a genuine fit. Leptons are measured best, jets not as well, while the missing transverse energy \MET has the largest uncertainty
In the lepton+jets final state one may, or may not, use \MET as the starting point for the transverse energy of the missing neutrino.
CDF and D0 use template, multivariate template, DLM, Matrix Element, ideogram, and multivariant discriminant analyses to select their top enriched and background samples of events that are basis of their top mass and cross section analyses. 

\begin{figure}[h]
\hspace{1.5cm}
\vspace{-0.5cm}
\includegraphics[width=0.79\textwidth]{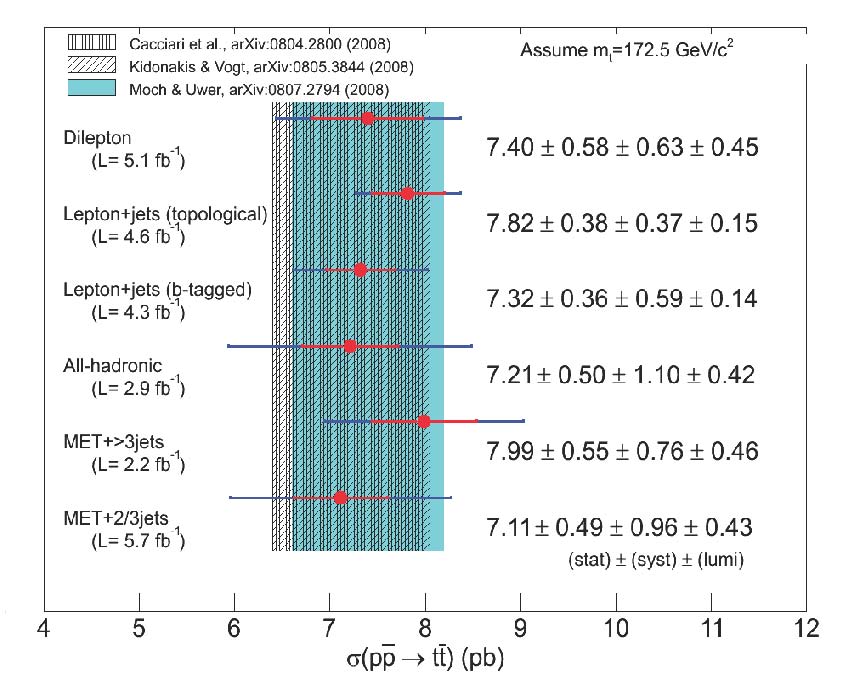}

\caption{\label{fig:TOPs} \small CDF $t{\bar t}$ cross section measurements in several final states, all analyses assume $M_{top}=172.5~GeV/c^2$.}
\end{figure}

\begin{figure}[h]
\hspace{1.5cm}
\vspace{0.0cm}
\includegraphics[width=0.79\textwidth]{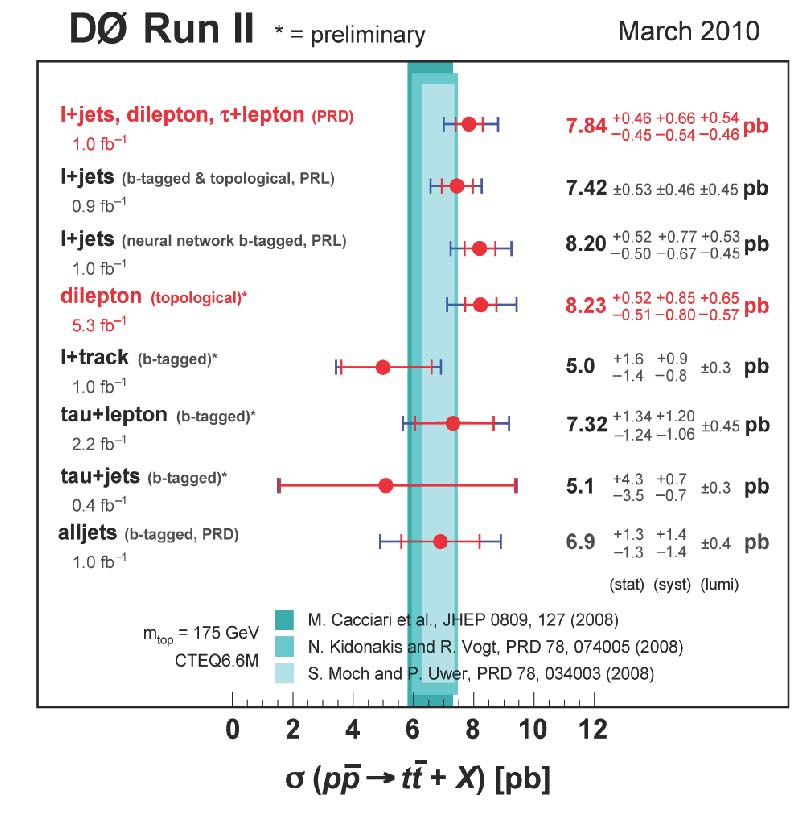}
\caption{\label{fig:TOPsD} \small D0 $t{\bar t}$ cross section measurements in several final states, assuming $M_{top}=175.0~GeV/c^2$. Predictions from various theoretical calculations are also shown on the plots.}
\end{figure}

\begin{figure}[ht]
\vspace{0.0cm}
\hspace{1.5cm}
\includegraphics[width=0.8\textwidth]{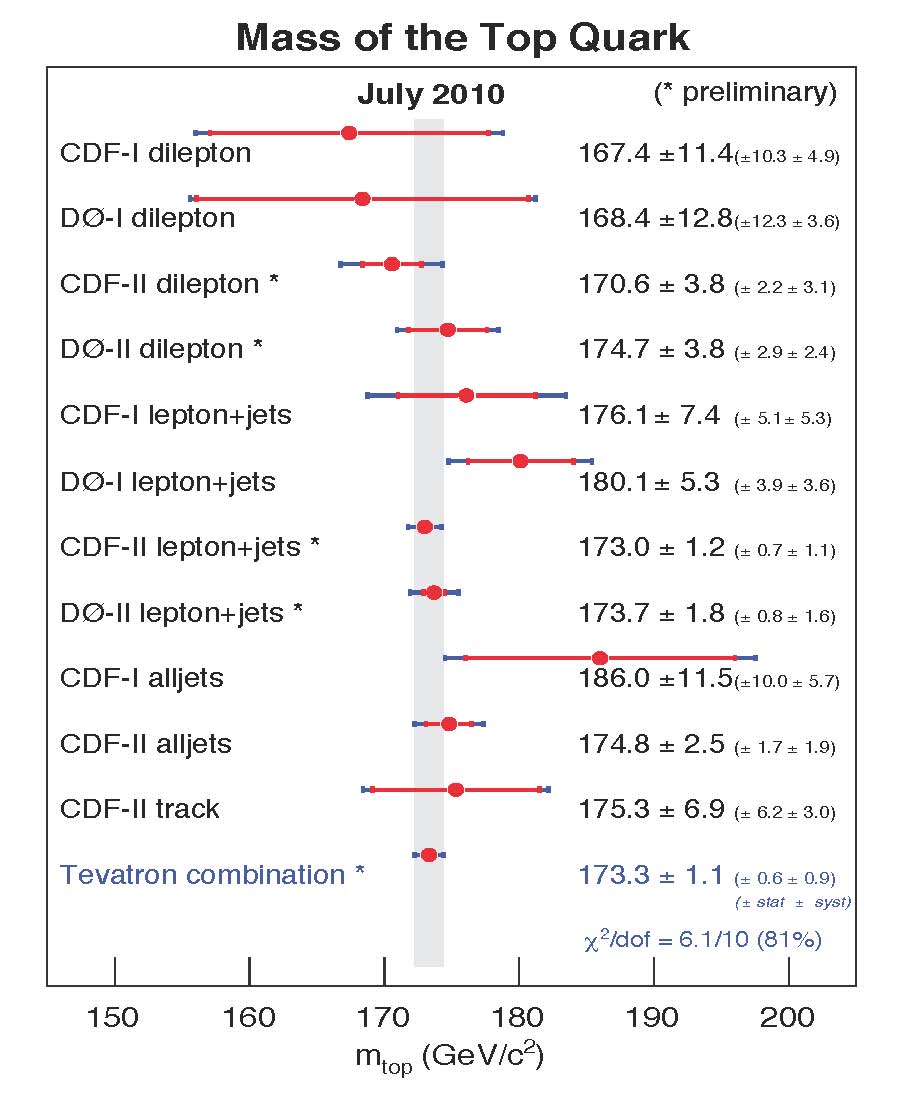}
\caption{\label{fig:TOPm} \small The error on combined measurement of $M_{top}$ is already at the value predicted for Tevatron with full dataset ($\Delta M_{top}\sim1.2~GeV/c^2$). (I remain a bit puzzled about the smallness of the systematic errors, I am worried that some effects may have yet not been taken into account.)}
\end{figure}

In the di-lepton mode situation is much more complicated, as the problem is under-constrained (two missing neutrinos). Several techniques were developed. All obtain a probability density distribution as a function of $M_t$ whose shape allows identifying the most likely mass which satisfies the hypothesis that a pair of top quarks were produced in an event and that their decay products correspond to a given combination of leptons and jets. \MET may, or may not, be used.
D0 and CDF developed several methods: the Neutrino Phase Space weighting technique ($\nu$WT); the Average Matrix Element technique (MWT); a modified form of Dalitz-Goldstein, Dalitz-Goldstein-Sliwa\cite{DG}  and Kondo\cite{DLM}  methods. Figures~\ref{fig:TOPs},\ref{fig:TOPsD} present the latest $t{\bar t}$ cross section measurements, and in Figure~\ref{fig:TOPm} a summary of the recent top quark mass measurement in different final states is shown. The combined D0 and CDF result for the top quark mass is\cite{TOPm}:
\begin{eqnarray*}
M_{top}& =& 173.3\pm0.6 (stat)\pm0.9 (syst)~GeV/c^2 
\end{eqnarray*}
Single top production, which is expected to be as small as some of the diboson channels, has been also observed by CDF and D0. In both experiments, analyses and searches used extensively NN and other advanced analysis techniques. The combined CDF result, based on 3.2/fb dataset is~\cite{STop} $2.3^{+0.6}_{-0.5}~pb$, and it is in good agreement with SM. The observation of single top adds to confidence that Higgs searches at Tevatron are feasible. 
\noindent
The future of top mass measurement is with the LHC experiments. LHC is a top factory. It will provide very large statistics very quickly - the challenge will be
to reduce the systematic errors.

\subsection*{Forward-backward asymmetry in top production}

Standard Model predicts a very small asymmetry, $A_{FB}^{theory}=0.05\pm0.015$ (based on NLO QCD calculations). CDF observed a much larger asymmetry for a long time\cite{Asym}, and recently D0 has confirmed this result\cite{D0Asym}.
\begin{eqnarray*}
CDF(5.3/fb):  A_{FB}& =& 0.158\pm 0.072 (stat)\pm 0.017 (syst)\\
D0 (4.3/fb) :  A_{FB}& =& 0.08\pm 0.04 (stat)\pm0.01 (syst)
\end{eqnarray*}
CDF has also looked at the dependence of the asymmetry as a function of the separation in rapidity $\vert \Delta y \vert = y_{t} - y_{\bar t}$, and found that most of the effect comes from $\vert \Delta y\vert > 1$. It is interesting to recall that in $e^+e^-$ colliders (SLAC, LEP), a corresponding asymmetry in b-quark production was also providing the most significant discrepancy with SM predictions. Recently, there has been increased theoretical activity to calculate $A_{FB}$ more reliably, however, those are not easy calculations (NNLO).  The result remains a puzzle. 

\subsection*{Higgs searches and exclusion limits}

\begin{figure}[h]
\vspace{-0.5cm}
\hspace{-1.5cm}
\includegraphics[width=1.2\textwidth]{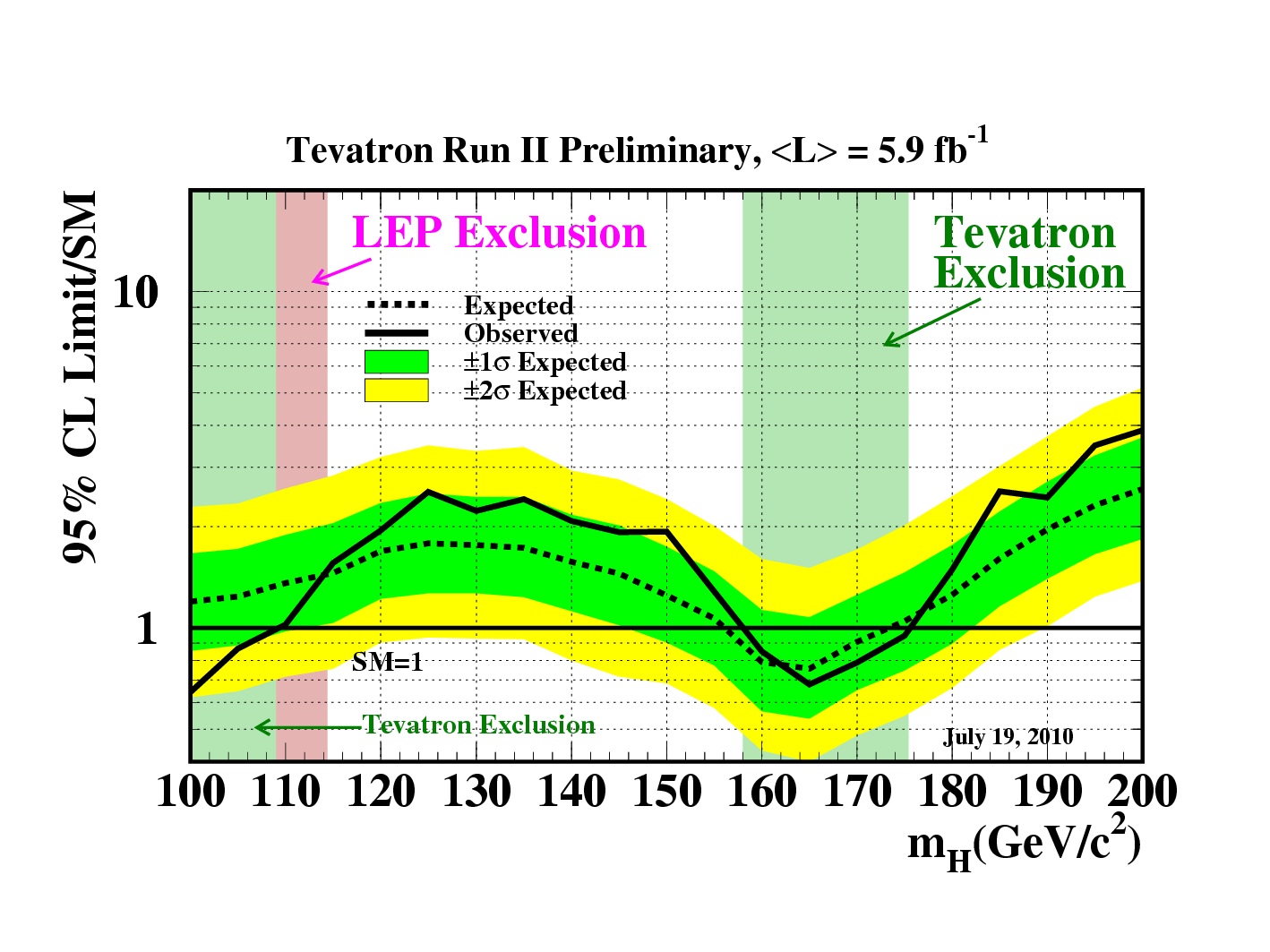}
\caption{\label{fig:Higgs} \small The combined CDF and D¯ upper Limits on Standard Model Higgs boson production based on 5.9/fb of data (summer 2010).}
\end{figure}

At Tevatron, according to MSM, Higgs can be produced in gluon fusion (largest cross section, but also large backgrounds), associated production with W,Z or a top pair (smaller cross section but cleaner) and vector boson fuson (even smaller signals, but may help improve sensitivity). In the low mass range ($M_H<135~GeV/c^2$) $b{\bar b}$ channel dominates, for higher masses  ($M_H>135~GeV/c^2$) all production modes are important, with the $WW$ and $WZ$ final states providing most of sensitivity. Both experiments use ALL accessible production modes and combine very small signals from a large number of final states and take advantage of advanced analysis techniques (NN, boosted decision trees, ME, etc). A Tevatron combination of all CDF and D0 analyses\cite{Higgs} (summer 2010), based on 5.9/fb of data, excludes the MSM Higgs in the range $(158<M_H<175)~GeV/^2$ at $95\%$ confidence level. The combined limit is shown in Figure~\ref{fig:Higgs}. The Tevatron projections indicate that  at the end of 2011, with 10/fb analysed data, the sensitivity of $>2.4\sigma$ will be reached in the Higgs mass range $(100<M_H<185)~GeV/c^2$. 

\section*{Tevatron future}

The same projections indicate that with luminosity of 16/fb the Tevatron sensitivity for MSM Higgs would reach $3\sigma$ in the range $(100<M_H<185)~GeV/^2$. This prediction was the basis of a proposal made to Department of Energy (DOE) in April 2010 to extend Run II by 3 years, through 2014, to reach 16/fb. At that time, the CERN plan was to oparate LHC at $\sqrt s =7~TeV$ until the end of 2011 and to collect 1/fb of data. A long (at least 1 year) shutdown to make LHC repairs and upgrades would follow, and LHC would re-start in 2013, capable of maximum energies and luminosities. In this scenario, extending Run II would keep Tevatron competitive in the MSM Higgs search. However, as a result of a very smooth LHC operation in 2010, a new CERN/LHC plan has been developed. In the new scenario the LHC shutdown will be postponed and the LHC experiments will continue to take data until the end of 2012, collecting 3-5/fb per experiment. The availability of such large amounts of LHC data already by the end of 2012, would make Tevatron running obsolete. At the time of my presentation at the Epiphany 2011, it was known that the proposal to extend Tevatron Run II has been given support from P5 advisory committee in October 2010, under the condition that new source of funding ($\sim35 M$\$/year) is identified to fund the Tevatron run extension. A few hours after my presentation, DOE announced its decision not to extend Tevatron Run II beyond September 2011. I announced this decision first thing on the next day at the Conference. The long and very successful Tevatron collider program is coming to its end. A few weeks later, after 2011 Epiphany Conference in Krakow had ended, at the yearly Chamonix meeting, the decision to extend LHC running through 2012 was indeed taken.

\section*{Acknowledgments}
\smallskip
I would like to thank the Conference
Organizers for their very warm hospitality and for making the Epiphany 2011 
an exceptionally well organized conference. It was also a very memorable experience for me to be invited to give an overview of the entire Tevatron collider program, with which I had been associated from its very beginning, on the same day that the decision to close it was made. 


\begin{thebibliography}{99}

\bibitem{SM} S.L. Glashow,\Journal{\NPB}{22}{579}{1961}; S. Weinberg, \Journal{\PRL}{19}{1264}{1967}; A. Salam, in Elementary Particle Physics (Nobel Symposium No.8), (ed. N. Svartholm). Almquist and Wilsell, Stockholm, 1968; F. Englert, R. Brout, \Journal{\PRL}{13}{321}{1964}; P.W. Higgs, \Journal{\PRL}{13}{508}{1964}; G. t'Hooft, \Journal{\NPB}{33}{173}{1971}, \Journal{\NPB}{35}{167}{1971}; G. t'Hooft, M.J.G. Veltman, \Journal{\NPB}{44}{189}{1972}.

\bibitem{det} R. Blair et al., The CDF II Detector Technical Design Report, Fermilab-Pub-96/390-E;  V. M. Abazov et al. (D0 Collaboration), Nucl. Instrum.
Methods Phys. Res., Sect. A 565, 463 (2006).

\bibitem{D0results}D0: http://www-d0.fnal.gov/Run2Physics/WWW/results.htm

\bibitem{CDFresults}CDF: http://www-cdf.fnal.gov/physics/physics.html

\bibitem{QCD}\Journal{\PRD}{80}{111107}{2009}

\bibitem{CDF_QCD}\Journal{\PRL}{88}{042001}{2002}

\bibitem{Wcomb} FERMILAB-TM-2439-E (arXiv:0908.1374 [hep-ex])

\bibitem{CP}\Journal{\PRL}{100}{161802}{2008}

\bibitem{D0asym}\Journal{\PRD}{82}{032001}{2010}

\bibitem{Sigma}\Journal{\PRL}{99}{202001}{2007}

\bibitem{Omega}\Journal{\PRL}{101}{232002}{2008}, \Journal{\PRD}{80}{ 072003}{2009}

\bibitem{spectr} CDF/PUB/BOTTOM/PUBLIC/10286

\bibitem{Y} \Journal{\PRL}{102}{242002}{2009}, CDF/DOC/BOTTOM/PUBLIC/10244

\bibitem{WW} \Journal{\PRD}{82}{031103}{2010}, CDF/PUB/ELECTROWEAK/PUBLIC/10238, FERMILAB-PUB-09-636 (arXiv: 0912.4500 [hep-ex]); arXiv: 1104.3078 [hep-ex].

\bibitem{Top}\Journal{\PRL}{73}{225}{1994}, F. Abe {\it et al}, \Journal{\PRL}{74}{2626}{1995}, S. Abachi {\it et al}, \Journal{\PRL}{74}{2632}{1995}.

\bibitem{DG}Gary R. Goldstein and R.H. Dalitz, \Journal{\PRD}{45}
{1531}{1992}; Gary R. Goldstein, K. Sliwa and R.H. Dalitz, \Journal{\PRD}
{47}{967}{1993}.

\bibitem{DLM}K. Kondo {\it et al.} J. Phys. Soc. Japan. {\bf 62}
(1993) 1177.

\bibitem{TOPm} arXiv:1007.3178 [hep-ex]

\bibitem{STop} \Journal{\PRL}{103}{092002}{2009}

\bibitem{Asym}\Journal{\PRL}{101}{202001}{2008}

\bibitem{D0Asym} D0 Note 6062-CONF

\bibitem{Higgs}arXiv:1007.4587 [hep-ex]

\end{thebibliography}
\end{document}